\documentclass[aps,prb,twocolumn,showpacs,superscriptaddress]{revtex4}
\usepackage{amsmath,amssymb}
\usepackage{graphicx}
\usepackage{times}

\begin{document}
\title{Probing vortex Majorana fermions and topology in semiconductor-superconductor heterostructures}
\author{Kristofer Bj\"{o}rnson}
\affiliation{Department of Physics and Astronomy, Uppsala University, Box 516, S-751 20 Uppsala, Sweden}
\author{Annica M. Black-Schaffer}
\affiliation{Department of Physics and Astronomy, Uppsala University, Box 516, S-751 20 Uppsala, Sweden}
\date{\today}

\begin{abstract}
We investigate the local density of states, spectral function, and superconducting pair amplitudes for signatures of Majorana fermions in vortex cores in ferromagnetic and spin-orbit coupled semiconductor-superconductor heterostructures.
We show that the Majorana fermion quasiparticle momentum distribution is always symmetrically distributed at a finite radius around a high symmetry point, thereby providing a necessary condition for a low-energy state to be a Majorana fermion. 
In real space profiles of the local density of states through the vortex core the Majorana fermion, together with other finite-energy vortex states, form a characteristic x-shape structure only present at non-trivial topology.
Moreover, we find that the Mexican hat band structure property of the topologically non-trivial phase translates into multiple high-intensity band edges and also vortex core states located above the superconducting gap in the local density of states.
Finally, we find no strong correlation between odd-frequency pairing and the appearance of Majorana fermions, but odd-frequency pairing exists as soon as ferromagnetism is present. In fact, we find that the only vortex superconducting pair amplitude directly related to any phase transition, is the appearance of certain spin-triplet $p$-wave pairing components in the vortex core at a pre-topological vortex core widening transition.
\end{abstract}
\pacs{74.90.+n, 03.65.Vf, 74.20.Rp, 74.25.Uv, 74.55.+v}

\maketitle

\section{Introduction}
In recent years the study of topological phases has lead to the prediction that so-called Majorana fermion quasiparticles can appear in certain types of topological superconductors.\cite{PhysRevB.61.10267, RevModPhys.82.3045, RevModPhys.83.1057, PhysUsp.44.131}
These Majorana fermions are of great interest for two quite distinct reasons.
First of all, Majorana fermions are hypothetical particles long sought in particle physics,\cite{NuovoCimento.5.171} but so far without any conclusive evidence in favor of their existence.\cite{NatPhys.5.614}
Majorana fermion quasiparticles in the solid state, although not fundamental particles, are analogous to their fundamental counter-parts, and may therefore provide an independent way to discover Majorana fermions.\cite{NatPhys.5.614}
The second reason they are of interest is because when they emerge in condensed matter systems, typically localized on various defects, they do so in a way which makes the ground state degenerate.
These different ground states can in two dimensions be continuously deformed into each other by braiding the defects around each other.
Such operations are predicted to be non-Abelian, and braiding Majorana fermions may thus be utilized for the implementation of robust topological quantum computing.\cite{PhysRevLett.86.268, RepProgPhys.75.076501}

One place where Majorana fermions are expected to emerge is in vortex cores in certain two-dimensional (2D) topological superconductors.
A prominent example of such a system is provided by a heterostructure of a thin layer of Rashba spin-orbit coupled semiconductor, sandwiched between a ferromagnet and a conventional $s$-wave superconductor.\cite{PhysRevB.77.220501, PhysRevLett.103.020401, PhysRevLett.104.040502, PhysRevB.82.134521}
Majorana fermions are also expected to occur at the end points of 1D wires of similar composition.\cite{PhysRevLett.105.077001, PhysRevLett.105.177002} Possible signatures of Majorana fermions has already been reported for such 1D systems,\cite{Science.336.1003, NatPhys.8.795, NatPhys.8.887, Science.346.6209} although some results are still debated.\cite{PhysRevB.90.085302, PhysRevB.91.094505}
Even if braiding in principle is possible for 1D systems through the use of wire networks,\cite{NatPhys.7.412} vortices lend themselves more naturally to be braided, for example through the use of magnetic force microscopy.\cite{RepProgPhys.75.076501, NaturePhysics.5.1169}
In the light of the potential versatility of vortex Majorana fermions it is of large interest to make a thorough investigation of different types of signatures for Majorana fermions in superconducting vortices.

Analytical and numerical results on vortices have already predicted that Majorana fermions appear in the topologically non-trivial phase of the above mentioned heterostructures.\cite{PhysRevB.77.220501, PhysRevLett.103.020401, PhysRevLett.104.040502, PhysRevB.82.134521, PhysRevB.88.024501}
In this work we carefully investigate experimentally relevant signatures of both the Majorana fermions and the topologically non-trivial phase.
In particular, we focus on signatures in the spectral function, local density of states (LDOS), and superconducting pair amplitude.
A simultaneous investigation of these three quantities is beneficial, as many signatures in one or another of these are closely related to particular features also in the other properties.

More specifically, we find that the Bogoliubov-de Gennes (BdG) quasiparticle spectrum for a Majorana fermion is strictly required to be symmetrically distributed at a finite radius around a high symmetry point, which thus provides a necessary condition for any candidate Majorana state. We also find that the vortex Majorana fermion and the generic finite-energy Caroli-Matricon-de Gennes vortex states\cite{PhysRevLett.9.307} are well separated in energy. Together they form a characteristic x-shape structure in subgap LDOS profiles right through the vortex core only in the topological phase. The Majorana mode is well-localized to the center of the core, while the finite-energy states disperse further out from the center.
Beyond the occurrence of Majorana fermions, the topologically non-trivial phase can also clearly be distinguished by a Mexican hat shaped band structure, which can be probed with momentum space probes such as angle-resolved photoemission spectroscopy (ARPES).
The Mexican hat shaped band structure further gives rise to multiple band edges, showing up as double peaks in the DOS.
The existence of the double peaks enables experimental techniques sensitive to the DOS to act as a probe of the topological phase, possibly most relevant for scanning tunneling spectroscopy (STS).
The two band edges in the double peak also behave differently in the presence of a vortex.
While one edge collapses to give rise to a Majorana fermion and the generic Caroli-Matricon-de Gennes subgap vortex core states, the other band edge is instead pushed up in energy and there gives rise to a second set of vortex core states that appears as a rising band edge in the LDOS spectrum of the vortex core.
We note that these types of STS measurable signatures are of particular interest, as the feasibility of such experiments recently have been demonstrated in a related setup consisting of an $s$-wave superconductor with vortices coated by a topological insulator.\cite{PhysRevLett.114.017001}
Finally, we show that there is no distinct onset of odd-frequency pairing as a result of the appearance of Majorana fermions, but odd-frequency pairing is present as soon as there is finite magnetism, independent on the topological phase.
However, we find a strong correlation between a pre-topological vortex core widening transition and the onset of $p$-wave pair amplitudes in and around the vortex core.

\section{Model}
We here consider a 2D topological superconductor with the essential building blocks being $s$-wave superconductivity, Rashba spin-orbit interaction, and Zeeman ferromagnetic term. To achieve self-consistent microscopic details for a vortex core we study the system on a square lattice.
The free parameters of the model are then the nearest neighbor hopping $t$ (setting the kinetic energy), chemical potential $\mu$, Zeeman field $V_z$, Rashba spin-orbit interaction $\alpha$, and superconducting pair potential $V_{sc}$.
All energies can be measured relative to the kinetic term, which we do by setting $t = 1$.
The Hamiltonian describing this system can be written as\cite{PhysRevLett.103.020401,PhysRevB.82.134521,PhysRevB.88.024501,PhysRevB.84.180509}
\begin{align}
\mathcal{H} &= \mathcal{H}_{kin} + \mathcal{H}_{V_z} + \mathcal{H}_{SO} + \mathcal{H}_{sc},
\label{Equation:Tight_binding_Hamiltonian} \\ 
\mathcal{H}_{kin} &= -t\sum_{\langle\mathbf{i},\mathbf{j}\rangle,\sigma}c_{\mathbf{i}\sigma}^{\dagger}c_{\mathbf{j}\sigma} - \mu\sum_{\mathbf{i},\sigma}c_{\mathbf{i}\sigma}^{\dagger}c_{\mathbf{i}\sigma}, \nonumber \\ 
\mathcal{H}_{V_z} &= -V_z\sum_{\mathbf{i},\sigma,\sigma'}(\sigma_z)_{\sigma\sigma'}c_{\mathbf{i}\sigma}^{\dagger}c_{\mathbf{i}\sigma'}, \nonumber \\ 
\mathcal{H}_{SO} &= -\frac{\alpha}{2}\sum_{\mathbf{i}}\left[ (c_{\mathbf{i}-\hat{x}\downarrow}^{\dagger}c_{\mathbf{i}\uparrow} - c_{\mathbf{i}+\hat{x}\downarrow}^{\dagger}c_{\mathbf{i}\uparrow})\right. \nonumber \\ 
&\textrm{\;\;\;\;\;\;\;\;\;\;\;\;\;\;\;\;}\left.+ i(c_{\mathbf{i}-\hat{y}\downarrow}^{\dagger}c_{\mathbf{i}\uparrow} - c_{\mathbf{i+\hat{y}}\downarrow}^{\dagger}c_{\mathbf{i}\uparrow}) + {\rm H.c.}\right], \nonumber \\ 
\mathcal{H}_{sc} &= \sum_{\mathbf{i}}\Delta_{\mathbf{i}}(c_{\mathbf{i}\uparrow}^{\dagger}c_{\mathbf{i}\downarrow}^{\dagger} + {\rm H.c.}) \nonumber.
\end{align}
Here $\mathbf{i}$ and $\mathbf{j}$ are site indices on the square lattice, $\sigma$ is the spin index, and $c_{\mathbf{i}\sigma}^{\dagger} (c_{\mathbf{i}\sigma})$ is the electronic creation (annihilation) operator.
We are primarily interested in a lightly hole-doped semiconductor, which is achieved by setting $\mu = 4$.
The superconducting order parameter $\Delta_{\mathbf{i}}$ enters as a parameter in the Hamiltonian, but is determined self-consistently using a superconducting pair-potential $V_{sc}$.
This is done by solving Eq.~\eqref{Equation:Tight_binding_Hamiltonian} within the BdG formalism, and re-calculating the order parameter using
\begin{align}
	\Delta_{\mathbf{i}}^{(m+1)}	=& -V_{sc}\langle c_{\mathbf{i}\downarrow}c_{\mathbf{i}\uparrow}\rangle^{(m)} \nonumber \\
								=& -V_{sc}\sum_{E_{\nu} < 0}v_{\nu\mathbf{i}\downarrow}^{(m)*}u_{\nu\mathbf{i}\uparrow}^{(m)}.
\end{align}
Here $u_{\nu\mathbf{i}\uparrow}$ ($v_{\nu\mathbf{i}\downarrow}$) is the electron up (hole down) component on site $\mathbf{i}$, and $m$ is the iteration step.
We are able to study a single vortex by specifying the initial order parameter configuration $\Delta^{(0)}(r,\theta) = |\Delta^{(0)}(r,\theta)|e^{-i\theta}$, but then letting the superconducting amplitude and phase fully relax.
This allows for a fully self-consistent order parameter profile to be obtained under the single requirement that the phase winds $2\pi$ around the vortex core.

\subsection{Phase transitions}
\begin{figure}
\includegraphics[width=245pt]{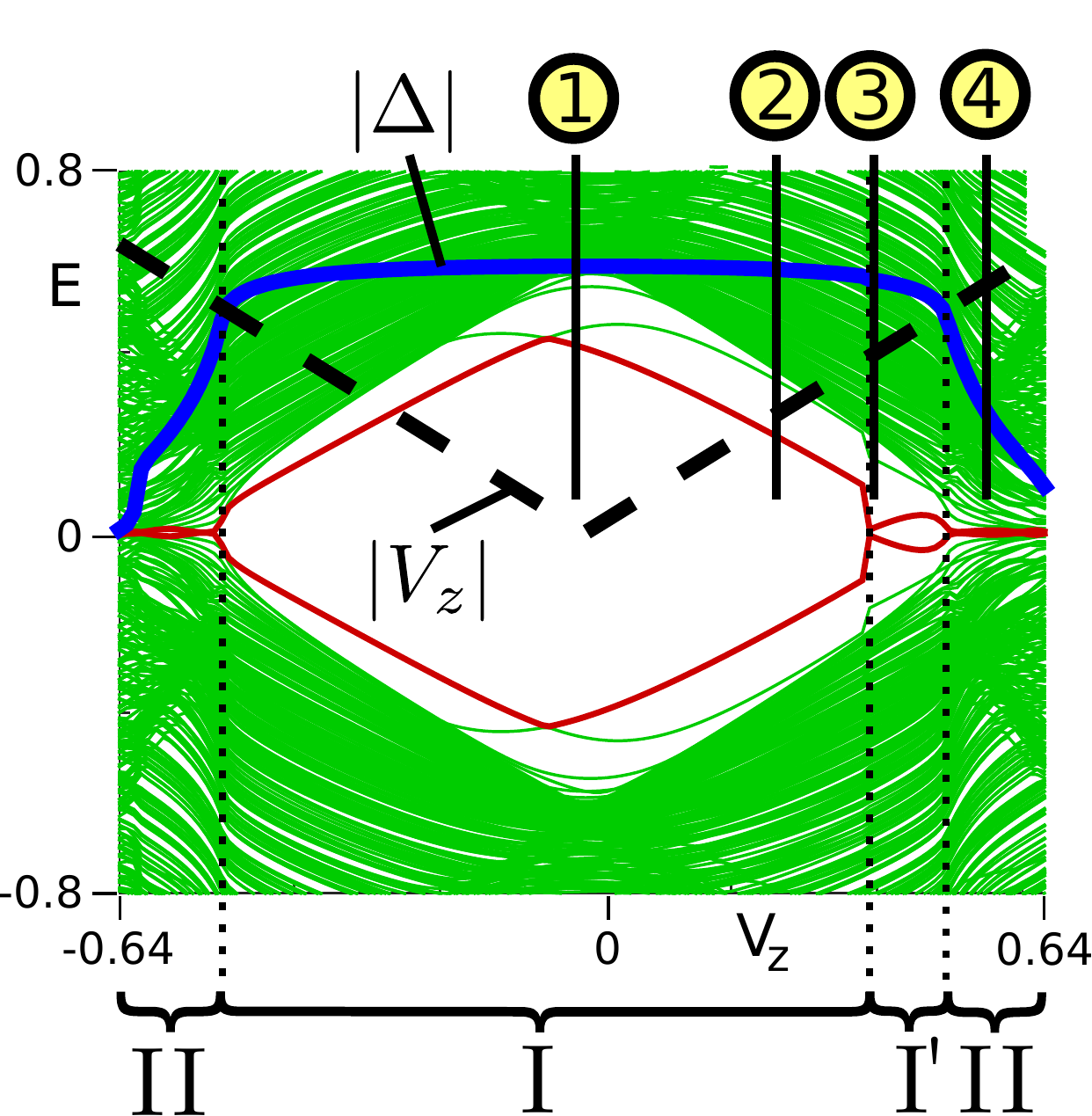}
\caption{(Color online). Energy spectrum for a superconducting vortex as a function of the Zeeman field (red and green lines), strength of the superconducting order parameter (thick blue line), and absolute value of the Zeeman field (black dashed line).
	The I, I', and II regions corresponds to the trivial phase, trivial phase with wide vortex core, and topologically non-trivial phase, respectively.
	Four representative sample points are also marked, used in other figures.}
\label{Figure:Energy_spectrum}
\end{figure}
We have previously shown that close to the topological phase an unrelated phase transition can take place, due to the competition between ferromagnetism and superconductivity.\cite{PhysRevB.88.024501}
This phase transition manifests itself in the sudden widening of the vortex core, together with a jump in the magnetization in the core.
This means that the phase diagram for the system can be divided into three different regions which we label I, I', and II.
These are the topologically trivial, topologically trivial but with a wide vortex core, and topologically non-trivial phases, respectively.
In Fig. \ref{Figure:Energy_spectrum} the energy spectrum as a function of the Zeeman ferromagnetic field is reproduced, and the three phases are marked with corresponding labels.
In the topologically non-trivial phase II, zero-energy Majorana states appear (red line).
However, states at or close to $E = 0$, which are not Majorana fermions, also appears in the I' region.
It is therefore directly clear that, if looking for a single signature such as a state at $E = 0$, there is a significant risk of mistakenly identifying a state as a Majorana fermion even though it is not.
In this work we therefore carefully investigate several different experimental signatures of both the Majorana fermions directly, as well as signatures related to the different phases.
Most of these signatures can be directly accessed with real space and band structure probes such as STS and ARPES, respectively, while a few other signatures are of at least important conceptual value.

\section{Bulk band structure}
\begin{figure*}
	\includegraphics[width=450pt]{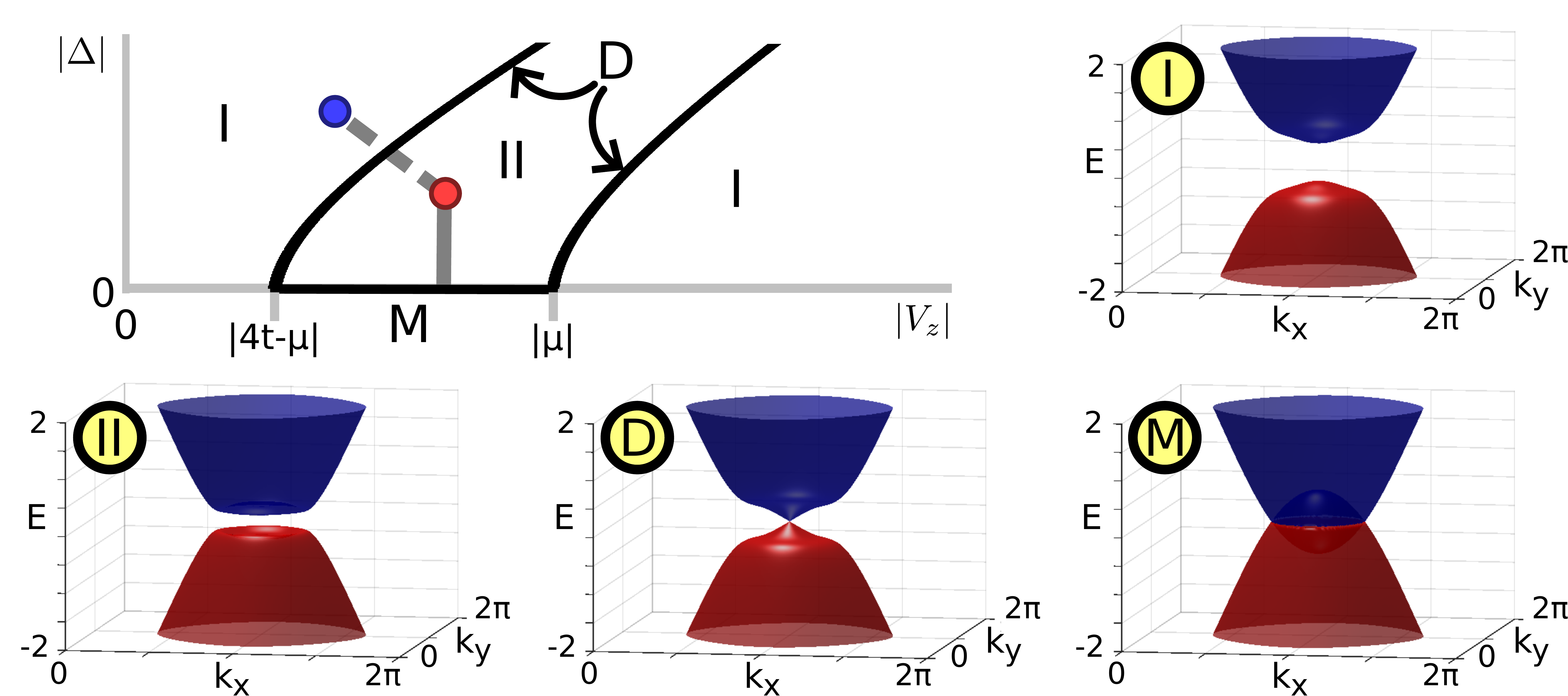}
	\caption{(Color online). [Top left] Schematic figure of the topologically trivial (I) and non-trivial (II) phases as functions of $|V_z|$ and $|\Delta|$. At the topological phase transition following the dashed grey line, the band gap is closed by the formation of a Dirac cone (D). The band gap can also be closed by following the full line down to $\Delta = 0$, at which point the system becomes metallic (M).
	[I, II, D, M] The two bands closest to the Fermi level plotted in the topologically trivial (I) and non-trivial phase (II), at the topological phase transition where a (D)irac cone appears, and for $\Delta = 0$ where the system becomes (M)etallic.}
	\label{Figure:Band_structure_gap_closing}
\end{figure*}

Before turning to the results for a vortex, we begin with a few important remarks about the bulk band structure and the topological phases of the system.
First of all, because we study a lightly hole-doped semiconductor, the relevant condition for being in the topologically non-trivial phase is\cite{PhysRevB.82.134521}
	$(4t - \mu)^2 + |\Delta|^2 < V_z^2 < \mu^2 + |\Delta|^2$.
We note that in the bulk there is no difference between the two topologically trivial phases I and I'.
The difference between these two phases only become apparent in self-consistent vortex calculations.
Further, the band structure of the system is given by\cite{PhysRevB.82.134521}
\begin{widetext}
	\begin{align}
	\label{Equation:Bulk_band_structure}
		E_{n}(\mathbf{k}) =& \pm\sqrt{\epsilon^2(\mathbf{k}) + \alpha^2\mathcal{L}_0^2(\mathbf{k}) + V_z^2 + |\Delta|^2 \pm 2\sqrt{\epsilon^2(\mathbf{k})\alpha^2\mathcal{L}_0^2(\mathbf{k}) + \left(\epsilon^2(\mathbf{k}) + |\Delta|^2\right)V_z^2}},
	\end{align}
\end{widetext}
where $n$ is the band index,
	$\epsilon(\mathbf{k})	= -2t\left(\cos(k_x) + \cos(k_y)\right) - \mu$ the kinetic energy, and
	$\mathcal{L}_0(\mathbf{k}) = \left(\sin(k_y), -\sin(k_x)\right)$ the spin-orbit coupling.
In Fig.~\ref{Figure:Band_structure_gap_closing}, we plot the band structure for representative points of the trivial and non-trivial phases, as well as points where the band gap closes.
It is clear from the plot for phase II that in the topologically non-trivial phase the two bands closest to $E = 0$ form two oppositely facing Mexican hats centered around the point $\mathbf{k} = (\pi,\pi)$.
[More generally this could also be at the other high symmetry point: e.g.~in a lightly electron-doped semiconductor the Mexican hats are centered around $\mathbf{k} = (0,0)$].
The trivial band structure in phase I instead takes the form of two ordinary parabolas.
Furthermore, the two plots (D) and (M) shows two different ways through which the bulk band gap of the non-trivial phase can be closed.
At the topological phase transition ($\textrm{I} \leftrightarrow \textrm{II}$), the band gap closes by forming a Dirac cone.
On the other hand, the band gap can also be closed by letting $\Delta \rightarrow 0$, in which case the system becomes metallic.
In the top left figure a schematic view of the two different routes through which the gap can be closed is shown.
From a topological point of view, the only important thing is that the I and II regions are separated by a gap closing, and the two types of gap closings can be considered topologically equivalent.
We will however see that by distinguishing between the two types of gap closings, the origin of important experimental signatures of Majorana fermions as well as the topological phase can be understood.

\subsection{Edge modes}
\label{Section:Extracting_information_from_the_bulk_band_structure}
In the semi-classical limit it is possible to treat real and reciprocal coordinates as independent of each other.
The edge states that appear at interfaces of topologically non-trivial phases can then be understood as a consequence of the bulk band structure being required to go through a gap closing when passing from the topologically non-trivial phase inside to the trivial phase outside.\cite{TheUniverseInAHeliumDroplet}
Zero-energy modes on impurities can be understood to appear for similar reasons.
However, in reality interface and impurities are fairly abrupt in nature, and it is therefore not obvious that a semi-classical treatment relying on the bulk band structure reliably predicts the properties of such defects.
Nevertheless, we know that the prediction of zero-energy states at interfaces is still valid also for abrupt defects, when the forced gap closing is associated with a jump in a topological invariant, such as the Chern number.
This have been demonstrated for many systems both numerically and analytically, and can in some cases be formally justified through the use of an appropriate index-theorem, see e.g.~\onlinecite{TheUniverseInAHeliumDroplet}.

\subsubsection{Semi-classical prediction}
In what follows we will repeatedly make use of the semi-classical limit to explain various signatures of Majorana fermions and the topological phase.
To demonstrate the method, we begin by determining the momentum space distribution of the Majorana fermions in the vortex core and at the edge of the system.
First, we note that the edge of a topologically non-trivial superconductor can be seen as an interface between the topologically trivial vacuum, and the non-trivial bulk.
It is therefore a transition from region I to II, similar to that along the dashed line in Fig.~\ref{Figure:Band_structure_gap_closing}.
In the semi-classical limit we thus expect the zero-energy spectrum on the edge to be related to a gap closing of type D, that is, the Majorana edge state will be located at the high symmetry point, $\mathbf{k} = (\pi, \pi)$.
On the other hand, the vortex core is a region where the most essential feature is that $\Delta \rightarrow 0$.
The Majorana state in the vortex core should therefore be expected to instead be related to a gap closing of type M, which produces zero-energy states in a circle at a certain radius from $\mathbf{k} = (\pi, \pi)$.

\subsubsection{Tight-binding confirmation}
\begin{figure}
\includegraphics[width=200pt]{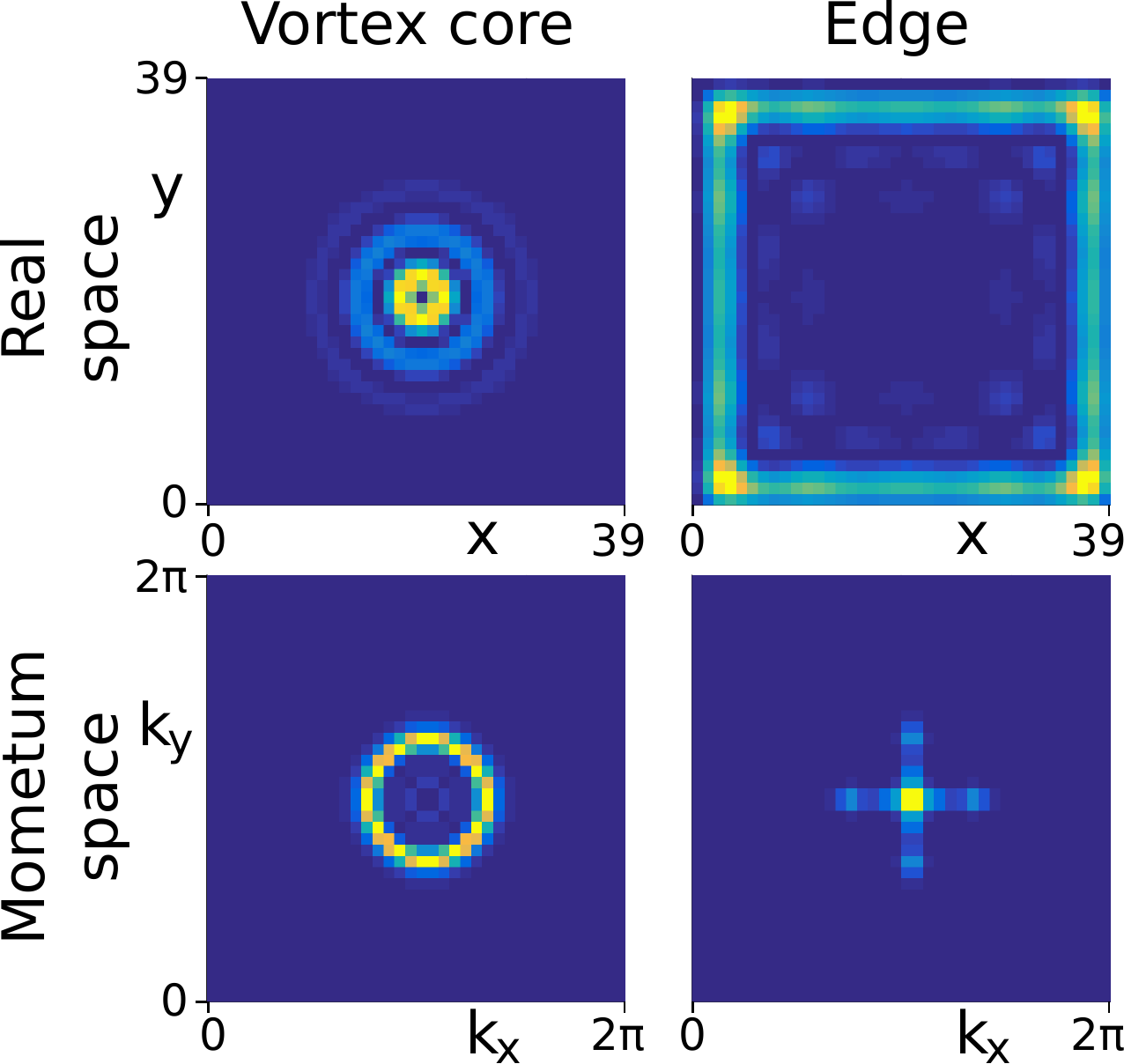}
\caption{(Color online). Majorana fermions in the vortex core and on the edge.
	Real space distribution (top) and Fourier transform of the same states (bottom).
	The vortex core Majorana fermion is mainly  built up of momentum components at a certain radius away from $\mathbf{k} = (\pi, \pi)$, while the Majorana state at the edge is consists predominantly of states at $\mathbf{k} = (\pi, \pi)$.}
\label{Figure:Majorana}
\end{figure}

Having deduced the expected behavior from the semi-classical limit, we now look at the actual Majorana fermions that appear in a fully self-consistent solution.
To do so we pick the eigenstates corresponding to the two zero-energy levels at point 4 in Fig.~\ref{Figure:Energy_spectrum}, and rotate these into the basis where the two resulting Majorana fermions $\gamma^{M_1}$ and $\gamma^{M_2}$ becomes clearly localized in the core and on the edge, respectively.
In Fig.~\ref{Figure:Majorana} we plot the square of the real space wave function given by
\begin{align}
	\label{Equation:Majorana_real_space}
	|\gamma^{M_{1,2}}(\mathbf{x})|^2 = \sum_{\sigma} |u_{\mathbf{x}\sigma}^{M_{1,2}}|^2 + |v_{\mathbf{x}\sigma}^{M_{1,2}}|^2.
\end{align}
We also we calculate the momentum space distribution of the states
\begin{align}
	\label{Equation:Majorana_momentum}
	|\gamma^{M_{1,2}}(\mathbf{k})|^2 = \sum_{\sigma} |u_{\mathbf{k}\sigma}^{M_{1,2}}|^2 + |v_{\mathbf{k}\sigma}^{M_{1,2}}|^2,
\end{align}
where $u_{\mathbf{k}\sigma}^{M_{1,2}}$ and $v_{\mathbf{k}\sigma}^{M_{1,2}}$ are the Fourier transforms of $u_{\mathbf{x}\sigma}^{M_{1,2}}$ and $v_{\mathbf{x}\sigma}^{M_{1,2}}$, respectively. The result is displayed in the reciprocal space plots in Fig. \ref{Figure:Majorana}.
It is clear that the Majorana fermion at the edge is mainly built up from $\mathbf{k}$-components at $\mathbf{k} = (\pi, \pi)$, while the vortex core Majorana fermion mainly consists of $\mathbf{k}$-components a finite radius away from $\mathbf{k} = (\pi, \pi)$.
This is in complete agreement with our expectations from the semi-classical limit.

Apart from showing that important features of the Majorana fermions can be derived directly from the bulk band structure by considering the semi-classical limit, this result also has important experimental implications;
a necessary indicator of the vortex core state being a Majorana fermion is that the Fourier transform of the state is distributed along a circle of finite radius, and centered at the high symmetry point where the bulk band gap closes at the topological phase transition.
However, we note that this is not an exclusive signature of a Majorana fermion, e.g.~ordinary Caroli-Matricon-de Gennes vortex states\cite{PhysRevLett.9.307} can be expected to have a similar momentum space distribution.
The reason being that also these states are a consequence of the same local collapse of the superconducting gap.

\section{Particle-hole symmetry}
\label{Section:MF_momentum_distribution}
Before moving on to further results, we here make a few general remarks related to particle-hole symmetry.
The BdG Hamiltonian for a general and homogeneous bulk superconductor can be written as a $4 \times 4$ matrix equation:
\begin{align}
	\mathcal{H}(\mathbf{k}) =& \left[\begin{array}{cc}
			\mathbf{H}_0(\mathbf{k})		& \boldsymbol{\Delta}(\mathbf{k})\\
			-\boldsymbol{\Delta}^{*}(-\mathbf{k}) & -\mathbf{H}_0^{T}(-\mathbf{k})
		\end{array}\right],
\end{align}
in the basis
	$(c_{\mathbf{k}\uparrow}, c_{\mathbf{k}\downarrow}, c_{-\mathbf{k}\uparrow}^{\dagger}, c_{-\mathbf{k}\downarrow}^{\dagger})^{T}$,
and we write the eigenstates as
$	\left|\gamma_{\nu\mathbf{k}}\right\rangle = \left[\begin{array}{cccc} u_{\nu\mathbf{k}\uparrow}	& u_{\nu\mathbf{k}\downarrow}	& v_{\nu\mathbf{k}\uparrow}	& v_{\nu\mathbf{k}\downarrow}\end{array}\right]^{T}$.
From the equation $\mathcal{H}(\mathbf{k})\left|\gamma\right\rangle = E\left|\gamma\right\rangle$ and the complex conjugate of $\mathcal{H}(-\mathbf{k})\left|\gamma\right\rangle = -E\left|\gamma\right\rangle$, it can be seen that these two equations are in fact the same set of equations.
This leads to a particle-hole symmetry for eigenstates $\nu$ and $\bar{\nu}$ of opposite energy $E_{\nu} = -E_{\bar{\nu}}$.
Identification of coefficients in the two equations also leads to
\begin{align}
	\label{Equation:Coefficient_relations_k}
	u_{\nu\mathbf{k}\sigma} = v_{\bar{\nu}-\mathbf{k}\sigma}^{*}.
\end{align}

Equation \eqref{Equation:Coefficient_relations_k} is of particular interest to us because it tells us that the states at $(E_{\nu}, \mathbf{k})$ and $(-E_{\nu}, -\mathbf{k})$ are related through
\begin{align}
	\label{Equation:Particle_hole_relation}
	\gamma_{\nu\mathbf{k}} = \sum_{\sigma}\left(u_{\nu\mathbf{k}\sigma}c_{\mathbf{k}\sigma} + v_{\nu\mathbf{k}\sigma}c_{-\mathbf{k}\sigma}^{\dagger}\right) = \gamma_{\bar{\nu}-\mathbf{k}}^{\dagger}.
\end{align}
This pairwise correspondence is a consequence of the artificial doubling of degrees of freedom which occurs when the problem is treated using a $4\times 4$ BdG formulation.
The two states are therefore not distinct, but rather the occupation of one is necessarily accompanied by the deoccupation of the other.

\subsection{Consequences for Majorana fermions}
The set of $\gamma_{\nu\mathbf{k}}$ provides a complete basis for the solution of any not homogeneous problem, with a general eigenstate written as
\begin{align}
	\label{Equation:General_BdG_quasi_particle}
	\gamma^{\lambda} = \sum_{\nu\mathbf{k}}a_{\nu\mathbf{k}}^{\lambda}\gamma_{\nu\mathbf{k}} = \sum_{\mathbf{k}\sigma}\left(u_{\mathbf{k}\sigma}^{\lambda}c_{\mathbf{k\sigma}} + v_{\mathbf{k}\sigma}^{\lambda}c_{-\mathbf{k}\sigma}^{\dagger}\right),
\end{align}
where $u_{\mathbf{k}\sigma}^{\lambda} = \sum_{\nu}a_{\nu\mathbf{k}}^{\lambda}u_{\nu\mathbf{k}\sigma}$ and $v_{\mathbf{k}\sigma}^{\lambda} = \sum_{\nu}a_{\nu\mathbf{k}}^{\lambda}v_{\nu\mathbf{k}\sigma}$.
Note that the eigenstates are now labeled by their superscript $\lambda$, while the subscript $\nu$ becomes a summation index.
If now $\gamma^{M}$ is a Majorana fermion, it also satisfies the relation $\gamma^{M} = \gamma^{M\dagger}$, which using Eqs.~\eqref{Equation:Particle_hole_relation}-\eqref{Equation:General_BdG_quasi_particle} leads to $\sum_{\nu\mathbf{k}}a_{\nu\mathbf{k}}^{M}\gamma_{\nu\mathbf{k}} = \sum_{\nu\mathbf{k}}a_{\bar{\nu}-\mathbf{k}}^{M*}\gamma_{\nu\mathbf{k}}$.
In particular, this implies
\begin{align}
	\label{Equation:Majorana_relation}
	u_{\mathbf{k}\sigma}^{M} = \sum_{\nu}a_{\nu\mathbf{k}}^{M}u_{\nu\mathbf{k}\sigma} = \sum_{\nu}a_{\bar{\nu}-\mathbf{k}}^{M*}u_{\nu\mathbf{k}\sigma} = v_{-\mathbf{k}\sigma}^{M*}.
\end{align}
It is therefore clear that the apparent symmetry of the momentum distribution in Fig.~\ref{Figure:Majorana} is not an accident.
Rather, Eq.~\eqref{Equation:Majorana_relation} guarantees that the momentum distribution of the Majorana fermion is inversion symmetric.
However, we note that both the real space and momentum distribution of the Majorana fermions expressed here takes into account both the electron and hole components of the eigenstates.
This strict symmetry is therefore not necessarily present for a physical probe only measuring the electron component.

\subsection{Electronic and BdG expressions}
The discrepancy just mentioned between what can be stated about the Majorana fermion in the BdG formulation and what can be seen experimentally reflect a general conceptual difficulty with regard to topological superconductors.
The topological properties are derived from the BdG band structure, so the consequences for these properties are most easily understood in relation to it.
However, physical quantities are calculated in a way that can obscure their relation to the BdG band structure.
A common prescription for calculating physical quantities such as the electronic LDOS, DOS, and spectral function involves summations of the form (see e.g.~Ref.~[\onlinecite{PhysRevB.78.024504}]):
\begin{align}
	\label{Equation:Physical_expressions_template}
	\sum_{E_{\nu} > 0}\left(|u_{\rho}^{\nu}|^2\delta(E - E_{\nu}) + |v_{\rho}^{\nu}|^2\delta(E + E_{\nu})\right),
\end{align}
where $\rho$ is some index and the summation runs over positive energies to avoid over-counting due to the artificial doubling of degrees of freedom in the BdG formulation.
We note that if the $\delta$-function had entered as $\delta(E - E_{\nu})$ in front of the $v$'s, then the $v$'s would appear as just another orbital, and such expressions can be compared straightforwardly with the BdG band structure.
For this reason it is beneficial to follow another equivalent prescription for calculating physical quantities.
Using Eq.~\eqref{Equation:Coefficient_relations_k}, or the same expression transformed to real space (depending on the type of index $\rho$): 
$	u_{\mathbf{x}\sigma}^{\nu} = v_{\mathbf{x}\sigma}^{\bar{\nu}*}$,
it is possible to replace terms involving $v$'s in expressions of the form in Eq.~\eqref{Equation:Physical_expressions_template}, by $u$ terms of the opposite energies.
We then write
\begin{align}
	\sum_{E_{\nu}}\left(|u_{\rho}^{\nu}|^2 + \left\{|v_{\rho}^{\nu}|^2\right\}\right)\delta(E - E_{\nu}),
\end{align}
where the second term in brackets is to be ignored in order for the expression to correspond to physical, or electronic, quantities.
On the other hand, if the second term is kept, the resulting properties can be straightforwardly related to the BdG band structure.
The latter expression, which we call BdG-type, therefore provides an important conceptual bridge between the theoretical formulation of (topological) superconductivity and experimentally measurable quantities. In fact, we already used the BdG-type expression in Eq.~\eqref{Equation:Majorana_real_space}-\eqref{Equation:Majorana_momentum} and Fig.~\ref{Figure:Majorana}.

\section{Spectral function}
\begin{figure}
\includegraphics[width=120pt]{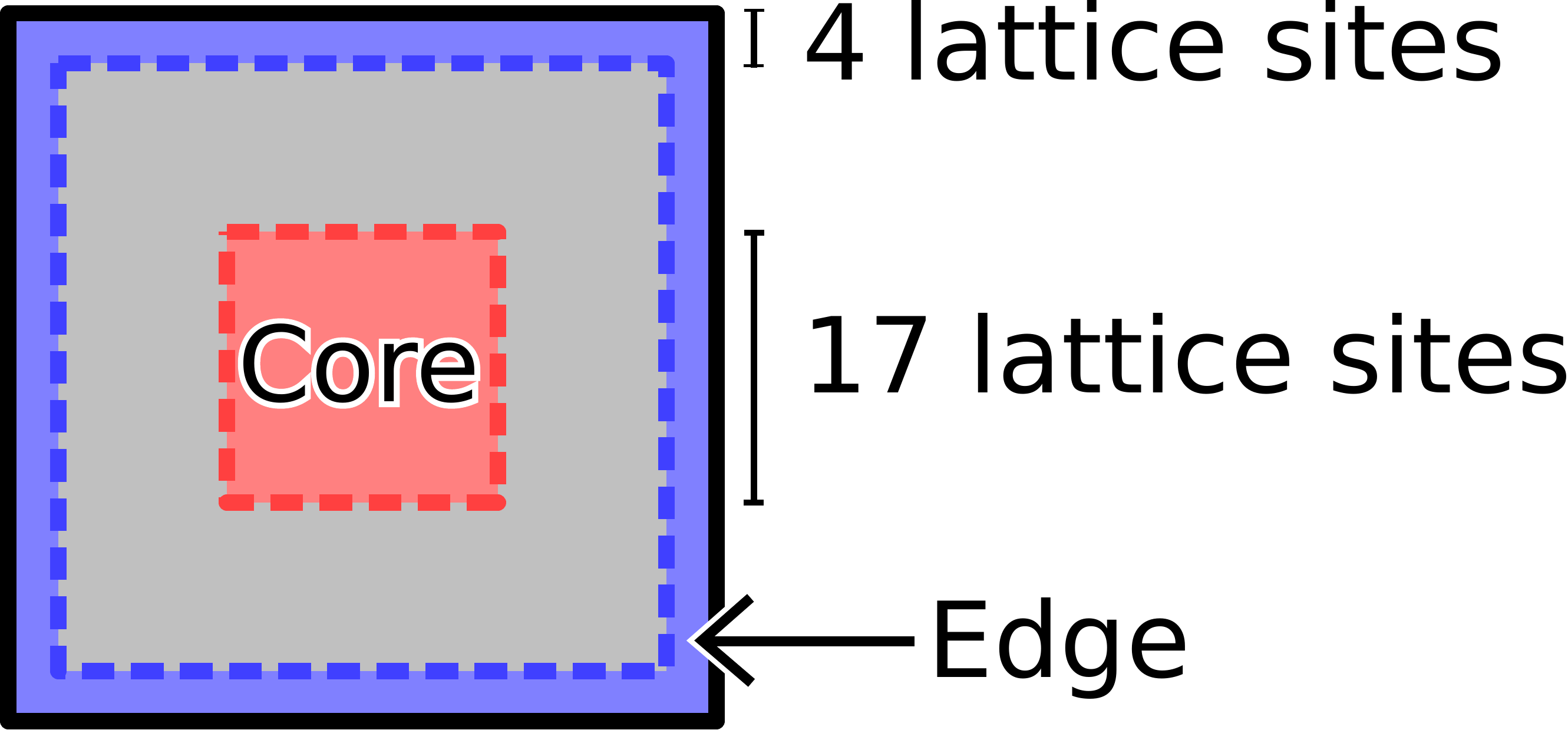}
	\caption{(Color online). Eigenstates are classified as edge, vortex core, or bulk states according to in which region they are mainly located. Eigenfunctions are classified as edge or vortex core states if more than 50\% of their density are located in the edge or vortex core region, respectively. Other states are classified as bulk states.
}
	\label{Figure:State_classification}
\end{figure}

In Section \ref{Section:Extracting_information_from_the_bulk_band_structure} we related the bulk band structure to the momentum space structure of the Majorana fermions.
Here we continue our study of the band structure by solving the real space Hamiltonian in Eq.~\eqref{Equation:Tight_binding_Hamiltonian} with a vortex, and then Fourier transform the results to arrive at the spectral function.
In the limit of an infinite homogeneous sample, this is equivalent to calculating the analytical bulk band structure.
However, when the sample is finite and includes defects such as edges and vortices, the result will be distorted.
The spectral function can be calculated as
\begin{align}
	A(\mathbf{k},E) =& \sum_{E_{\nu},\sigma}\left(|u_{\nu\mathbf{k}\sigma}|^2 + \left\{|v_{\nu\mathbf{k}\sigma}|^2\right\}\right)\delta(E - E_{\nu}).
\end{align}
It is useful to divide $A(\mathbf{k}, E)$ into parts consisting of edge, vortex core, and bulk contributions.
For this reason we define the state classification functions
\begin{align}
	\label{Equation:State_classification}
	C_{e}(\nu) =& \left\{\begin{array}{cc}
					1	&	\textrm{if\;} \sum_{x\in X_e,\sigma}\left(|u_{\nu\mathbf{x}\sigma}|^2 + |v_{\nu\mathbf{x}\sigma}|^2\right) > \frac{1}{2},\nonumber \\
					0	&	\textrm{otherwise},
				\end{array}\right.\\
	C_{c}(\nu) =& \left\{\begin{array}{cc}
					1	&	\textrm{if\;} \sum_{x\in X_c,\sigma}\left(|u_{\nu\mathbf{x}\sigma}|^2 + |v_{\nu\mathbf{x}\sigma}|^2\right) > \frac{1}{2},\nonumber \\
					0	&	\textrm{otherwise},
				\end{array}\right.\\
	C_{b}(\nu) =& \left\{\begin{array}{cc}
					1	&	\textrm{if\;} C_{e} = C_{c} = 0,\\
					0	&	\textrm{otherwise},
				\end{array}\right.
\end{align}
where $X_e$ and $X_c$ are the set of points classified as edge and core sites according to Fig. \ref{Figure:State_classification}.
Letting $L=e,c,b$ denote edge, core, and bulk, respectively, we define
\begin{align}
	\label{Equation:Dynamic_structure_factor_components}
	A_{L}(\mathbf{k},E) =& \sum_{E_{\nu},\sigma}C_{L}(\nu)\left(|u_{\nu\mathbf{k}\sigma}|^2 + \left\{|v_{\nu\mathbf{k}\sigma}|^2\right\}\right)\delta(E - E_{\nu}), 
\end{align}
where $A(\mathbf{k}, E) = \sum_L A_L(\mathbf{k}, E)$.
A peculiarity of this is that the numerical zero-energy states, representing the Majorana fermions before they are rotated into the Majorana basis, are classified as bulk states.
This follows since they are located both at the edge and in the vortex core, with additional tails stretching into the bulk.
This conveniently allows us to extract from $A(\mathbf{k},E)$ the contribution from the bulk band and Majorana fermions separately from the contributions from edge states and other vortex core states.

\begin{figure}
	\includegraphics[width=245pt]{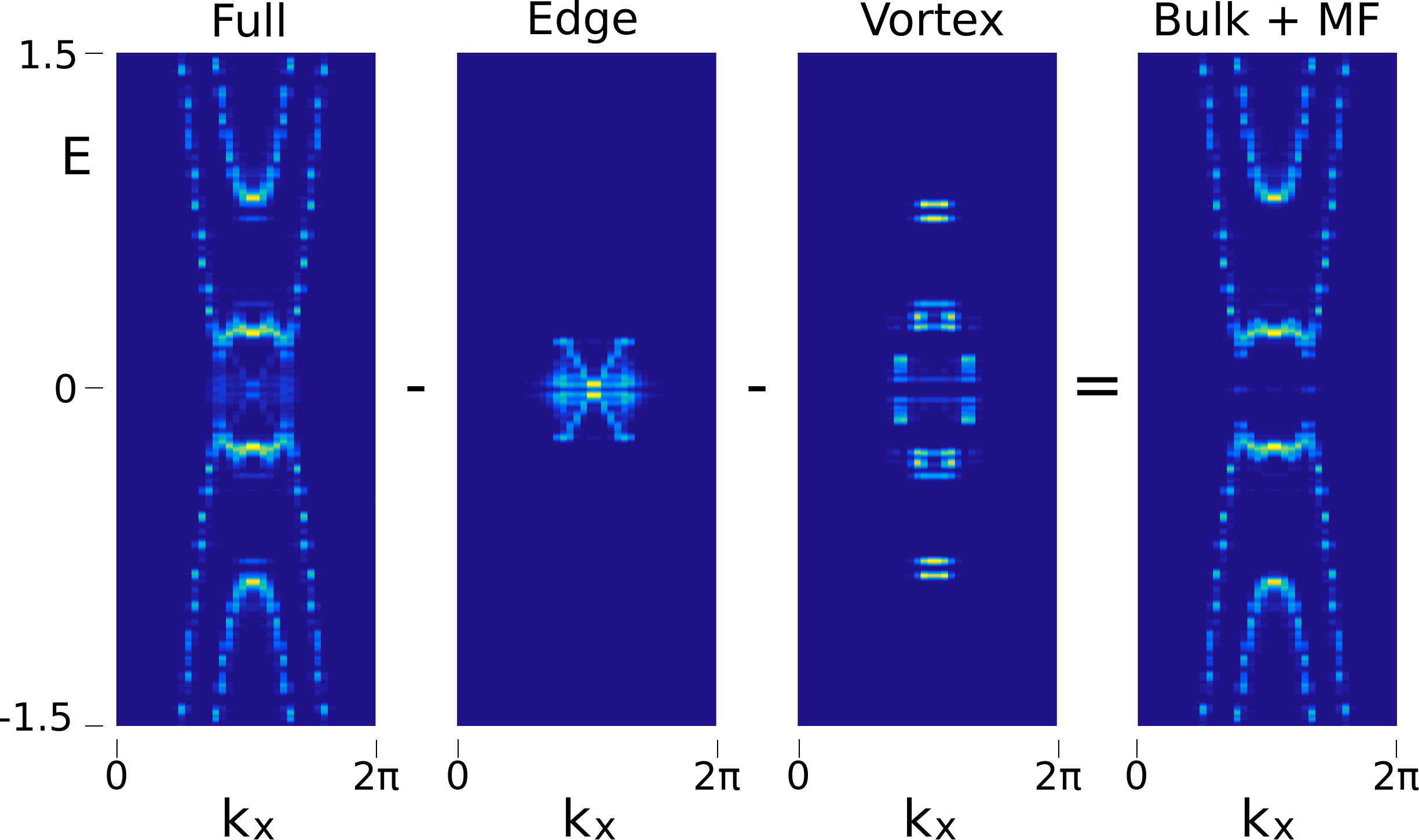}
	\caption{(Color online). The  BdG-type spectral function $A(\mathbf{k}, E)$ and its edge, core, and bulk components plotted along $k_y = \pi$ for parameters corresponding to the topologically non-trivial phase at point 4 in Fig.~\ref{Figure:Energy_spectrum}.
$A(\mathbf{k}, E)$ is seen to be dominated by contributions from the bulk band structure.
	However, topological edge states, as well as vortex core states also contribute.
	When the edge and ordinary vortex core states are subtracted from $A(\mathbf{k}, E)$, the resulting $A_b(\mathbf{k}, E)$ consists of only bulk bands and the two zero-energy Majorana fermion states.
	Because the scaling behavior is different for the different types of states, the color scale is arbitrary but set to enhance the relevant features.
	}
	\label{Figure:Band_structure_decomposition}
\end{figure}

In Fig.~\ref{Figure:Band_structure_decomposition} we plot the BdG-type $A(\mathbf{k}, E)$ and $A_{L}(\mathbf{k}, E)$ along $k_y = \pi$, at the topologically non-trivial point 4 in Fig. \ref{Figure:Energy_spectrum}.
First of all, we see the two dispersive (at non-zero energy) branches of topological edge states in $A_e(\mathbf{k}, E)$.
Similarly, a wealth of vortex core states are visible in $A_c(\mathbf{k}, E)$.
We note that vortex core states predominantly originate from states around the band edges.
This can be understood by considering that at the band edges the DOS is high, which makes it easy to hybridize these states into states localized in the vicinity of the vortex.
We also see that the ordinary vortex core states closest to $E = 0$ are formed mainly from $\mathbf{k}$-components at a distance away from $\mathbf{k} = (\pi, \pi)$, while higher energy vortex states are located notably closer to $\mathbf{k} = (\pi, \pi)$.
The former, which are subgap states, are the ordinary Caroli-Matricon-de Gennes states, and their momentum distribution is in agreement with that expected from our discussion in Section \ref{Section:Extracting_information_from_the_bulk_band_structure}, where we mentioned that they, just like the vortex core Majorana fermions, result from a local (M)-type collapse of the bulk gap.
In addition, it is clear that $A_b(\mathbf{k}, E)$ contains the bulk band structure, as well as faint signals of the Majorana fermions at $E = 0$.

Although it is possible to image the electronic $A(\mathbf{k}, E)$ directly using e.g.~ARPES, it may be difficult to extract the edge and vortex core features from such data, due to limited intensity.
However, the bulk features such as the Mexican hat shaped band structure in Fig.~\ref{Figure:Band_structure_decomposition} scale with the area of the bulk, and can therefore be imaged directly with ARPES.
On the other hand, it should be feasible to image the vortex core contribution by Fourier transforming data from local probes such as STS.
Note, however, that it is not the BdG-type spectral function in Fig.~\ref{Figure:Band_structure_decomposition} that is physically measured.
Rather it is the corresponding electronic spectrum which results from dropping the hole-part in Eq.~\eqref{Equation:Dynamic_structure_factor_components}.
%
Having seen how various features of the spectral function can be understood by splitting it decomposing into edge, vortex core, and bulk states contributions, we define
$	A_{b+c}(\mathbf{k}, E) = A_b(\mathbf{k}, E) + A_c(\mathbf{k}, E)$.
This is the spectral function with contributions from edge states excluded, which are artificial effects introduced by the finites size of our sample.
We will use this spectral function in the rest of this article.

\section{Local density of states}
\begin{figure*}
	\includegraphics[width=490pt]{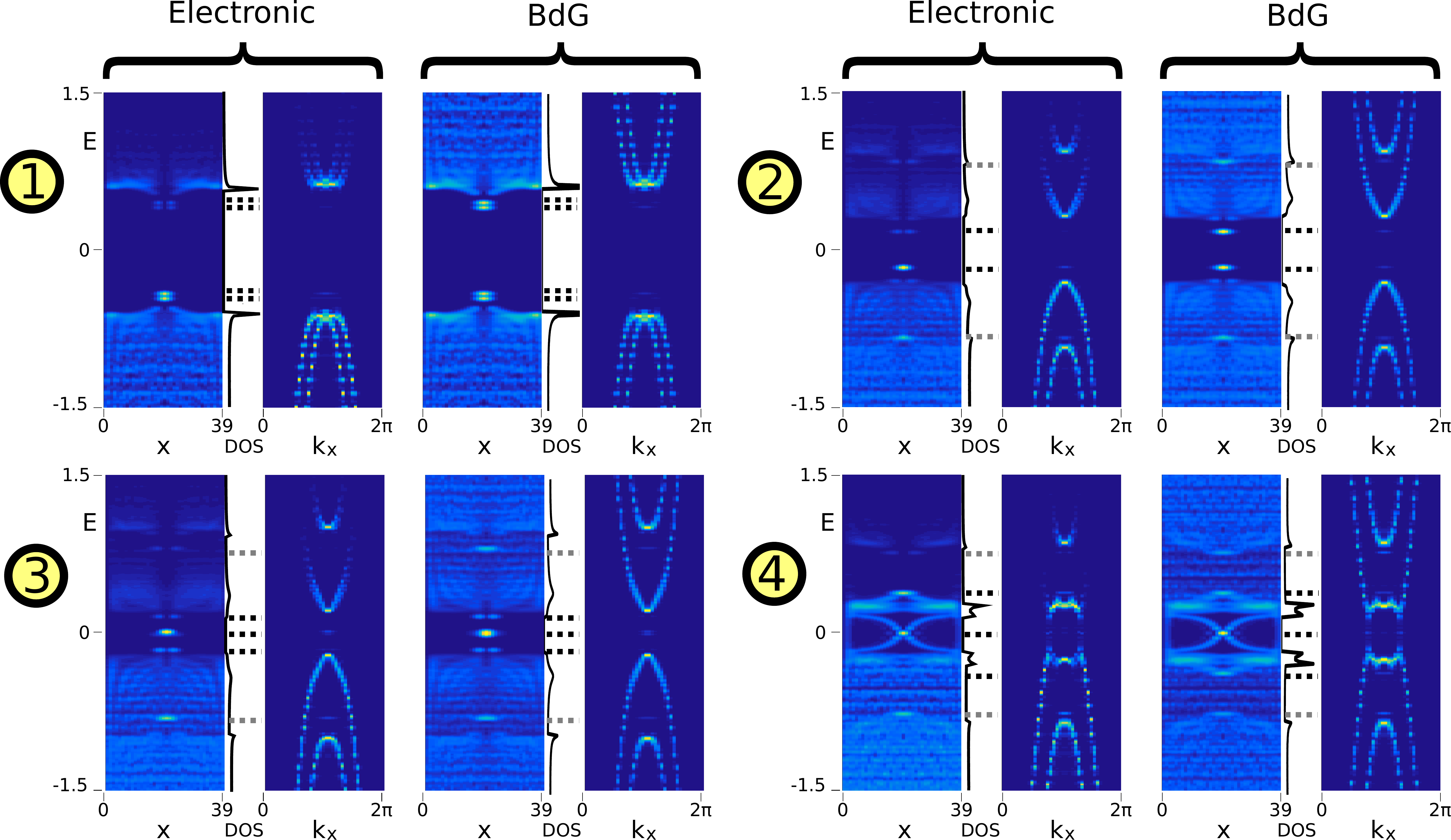}
	\caption{(Color online). LDOS for a cut through the vortex core (left), bulk DOS (center), and spectral function at $k_y = \pi$(right), for the four sample points shown in Fig.~\ref{Figure:Energy_spectrum}, using both the electronic and BdG-type expressions.
	The position of the vortex core states in the LDOS and spectral function are indicated in the DOS plots by dotted lines, and appears in grey if they are associated with the two higher-energy bands.
	}
	\label{Figure:LDOS_DSF}
\end{figure*}

Having investigated signatures in momentum space, we now turn to a discussion of the LDOS, given by
\begin{align}
	\rho(\mathbf{x}, E) = \sum_{\nu,\sigma}\left(|u_{\nu\mathbf{x}\sigma}|^2 + \left\{|v_{\nu\mathbf{x}\sigma}|^2\right\}\right)\delta(E - E_{\nu}).
\end{align}
We will see that a comparison between the LDOS and spectral function is helpful for revealing important signatures of the topological phase.
In addition we also compare with the bulk DOS calculated from a pure bulk solution with an equivalent $|\Delta|$.
%
In Fig. \ref{Figure:LDOS_DSF} the LDOS, bulk DOS, and $A_{b+c}(\mathbf{k}, E)$ are plotted side by side at the points labelled 1-4 in Fig. \ref{Figure:Energy_spectrum}.
Both the electronic and BdG-type results are shown here.
In what follows we use the BdG-type results for the interpretation, but the electronic results are to be considered when comparing to experiments.
When the Zeeman field is zero (1), the band structure can be seen to consist of two Rashba spin-orbit split parabolas crossing at a high symmetry point.
The band edges shows up as sharp features in both the LDOS and DOS.
Vortex core states below the band edges can also be observed.
As the Zeeman field is turned on (2), the two bands are split off from each other, and the upper band becomes irrelevant for the low-energy spectrum making the system effectively "spinless".
Still vortex core states form below both band edges.
Inside the I' region (3), a vortex core state appears at or close to $E = 0$, and there gives rise to a non-Majorana fermion zero bias peak.
It is clear from the spectral function that the topologically non-trivial phase has not yet been entered, as the two parabolic bands closest to the Fermi level do not give a Mexican hat shaped spectrum.
However, this can not be seen from the LDOS at $E = 0$ alone, and care thus has to be taken to not casually interpret every zero-energy peak in the LDOS spectrum as a Majorana fermion.
We further note that all vortex core states in these first three cases are associated with faint signals in the spectral function around the high-symmetry point and below the band edges.
This is particularly important for the zero-energy states in region I', as this make their momentum distribution clearly distinct from those of the Majorana vortex core states.

Once the two low-lying bands becomes inverted and acquires Mexican hat like forms, the topologically non-trivial phase is entered, as exemplified by point (4).
Because of the comparatively rather flat nature of the Mexican hat like structure, the band edges acquires a large density of states, which generates a strong intensity in the (L)DOS.
The observation of strong band edges in the LDOS spectrum therefore provides evidence for the system being in the topologically non-trivial phase.
However, this only distinguishes the topologically non-trivial phase from the trivial phase, and not from a conventional superconductor where strong band edges also appears.
We also note that the bulk DOS clearly shows that the band edge in fact consists of two edges, one from the Mexican hats lower edge away from $\mathbf{k} = (\pi, \pi)$, and one from its edge at higher energy at $\mathbf{k} = (\pi, \pi)$.
This provide a clear distinction also from the band edge behavior of a conventional superconductor.
Additionally, we see that the zero-energy Majorana fermion is accompanied by a wealth of other vortex states, showing up as a distinct x-shape structure in the low-energy LDOS.
These subgap states are the ordinary Caroli-Matricon-de Gennes states already discussed.

\begin{figure}
	\includegraphics[width=245pt]{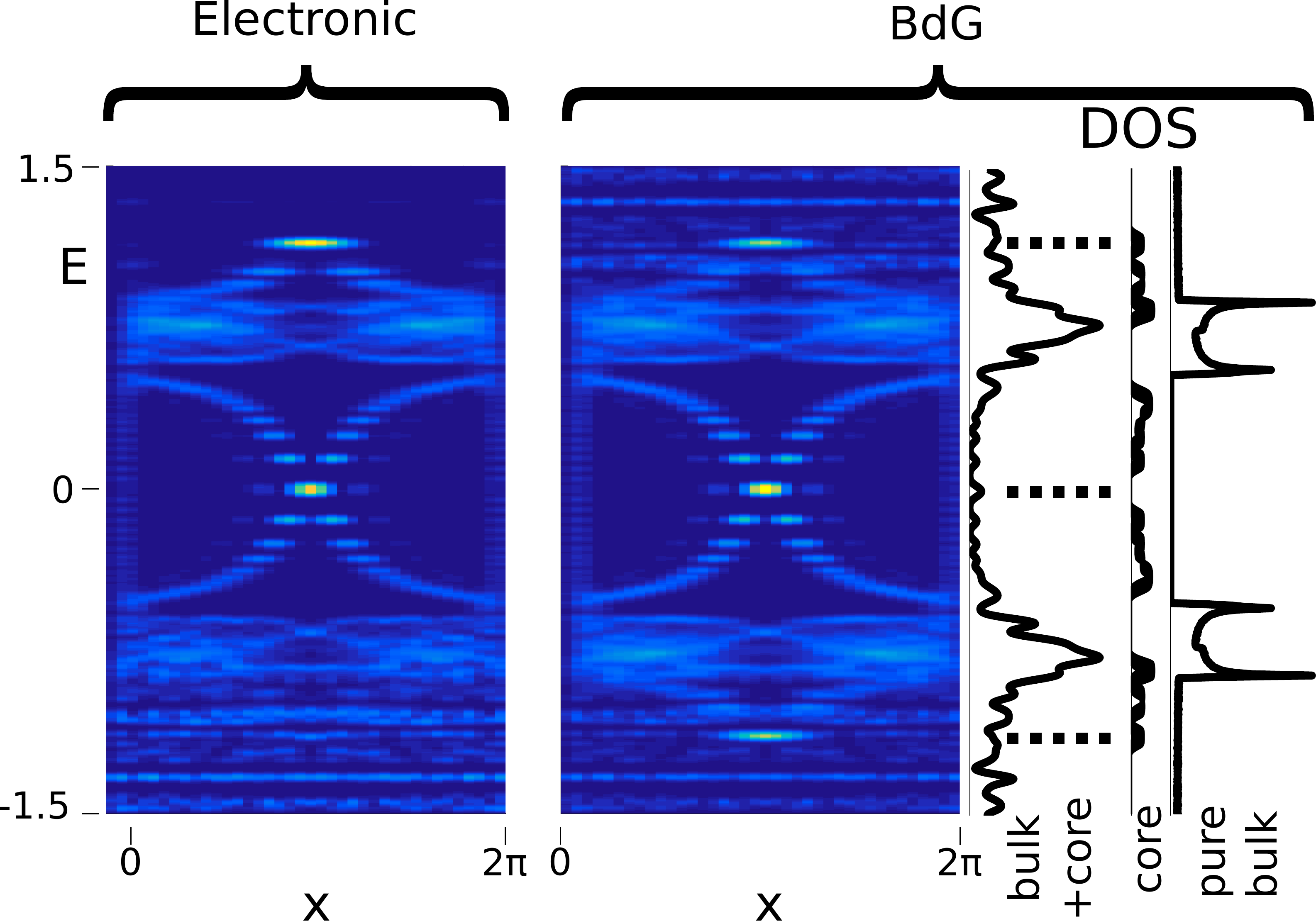}
	\caption{(Color online). Same LDOS as in (4) in Fig. \ref{Figure:LDOS_DSF}, but with higher resolution around $E = 0$. The vortex core states are now seen to be clearly separated by finite energies. The DOS is plotted to the right. Leftmost DOS consists of bulk and core contributions from vortex calculation. The middle DOS contains only vortex core states. Rightmost DOS is from a pure bulk calculation.}
	\label{Figure:LDOS}
\end{figure}

Another important piece of evidence for the topologically non-trivial phase is provided by an apparently larger band gap inside the vortex core.
From the LDOS it appears as if the band gap is increased around the vortex, leaving a strong vortex core signal above the bulk band edge.
This is due to the fact that the vortex core states now behave differently than the vortex core states observed so far, which always appear below the bulk band edge.
We can understand this as a consequence of the two different band edges of the Mexican hat shaped bulk band.
The lower band edge away from $\mathbf{k} = (\pi, \pi)$ collapses in the core because it is due to superconductivity.
However, the second band edge does not collapse.
Rather it can be seen from Eq.~\eqref{Equation:Bulk_band_structure} that at $\mathbf{k} = (\pi,\pi)$ the energy is given by $E_n = \pm\left(|V_z|-\sqrt{(4t-\mu)^2 + |\Delta|^2}\right)$.
It is further clear that in the topologically non-trivial phase the square root term is necessarily smaller than $|V_z|$.
Thus, as $|\Delta| \rightarrow 0$ inside the vortex core, the second band edge is pushed away from the Fermi level.
This result in a set of vortex core states above the band edge.
For more details of this process we provide a zoomed-in plot of the LDOS at point (4) in Fig.~\ref{Figure:LDOS}.
First of all, the Majorana fermion is clearly visible as a state at $E = 0$, and has a clear energy-separating from the other Caroli-Matricon-de Gennes states, making it possible to resolve.
Shown is also the BdG-type DOS, obtained by summing the bulk and vortex core DOS of the self-consistent solution, the vortex core DOS for the self-consistent solution, as well as the DOS of the pure bulk solution.
A comparison between the self-consistent bulk+core DOS and the DOS of the pure bulk solution clearly show how the lower band edge essentially collapses around the vortex and gives rise to sub gap states.
On the other hand, the upper band edge is pushed up in energy, giving rise to the vortex core states above the band edge, most clearly visible as a half x-shape structure above the band edge in the electronic spectrum.

We finally also point out that the Majorana fermion is well localized in the center of the vortex core, and that the arms of the x-shape structure formed by the subgap vortex core states meet at the center of the vortex.
We put this in contrast to recent STS experiments, where a spatially extended zero-energy peak, together with the arms of the x-shape structure pointing towards zero a finite distance away from the center of the vortex, were assumed to provide evidence of Majorana fermions in a superconductor-topological insulator heterostructure.\cite{PhysRevLett.114.017001}
Our results clearly show that Majorana fermions do not produce such signatures in a spin-orbit coupled semiconductor.
However, it should be noted that these experiments were done on an $s$-wave superconductor coated by a topological insulator, and may therefore be sufficiently different that a direct comparison is not possible.

\section{Pair amplitudes}
\begin{figure}
	\includegraphics[width=245pt]{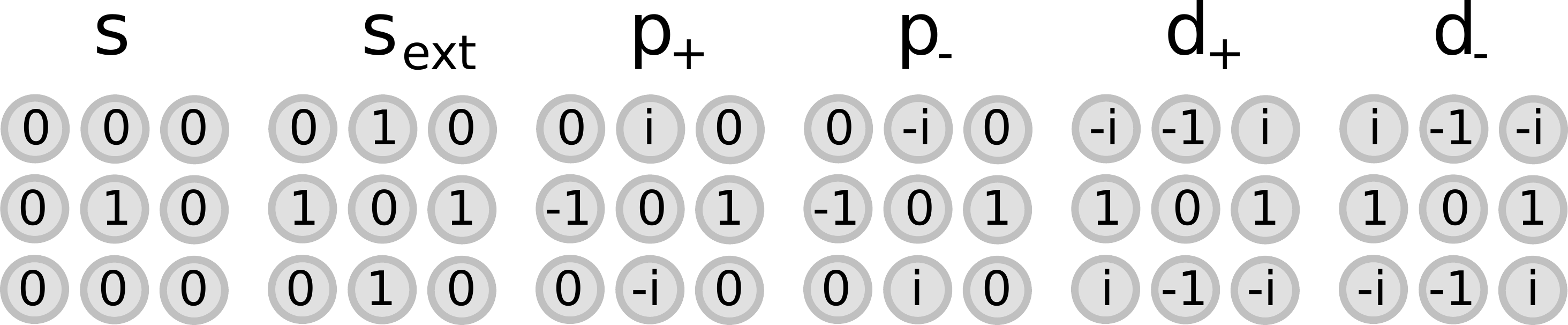}
	\caption{Orbital wave functions defined by the projection operators in Eq.~\eqref{Equation:Projectors}.
	The values on the sites are the complex conjugate of the value of the projector when $\mathbf{r}$ points from the central site to the corresponding site.}
	\label{Figure:Projections}
\end{figure}

It has recently been suggested that the appearance of Majorana fermions is closely related to the presence of odd-frequency pairing.\cite{PhysRevB.86.064512, arXiv.1410.1245}
In addition, the appearance of unconventional pair amplitudes in vortices has attracted great interest in general.\cite{JPhysCSolidStatePhys.21.L215, JPhysCondensMatter.1.277, NewJPhys.11.075008, PhysRevB.84.064530, PhysRevB.88.104506}
For this reason we here also provide a detailed investigation of the superconducting pair amplitude
\begin{align}
	\label{Equation:Pair_function}
	F(\mathbf{R}, \mathbf{r}, \sigma, \sigma') =& \langle c_{\mathbf{R}+\mathbf{r},\sigma}c_{\mathbf{R}-\mathbf{r},\sigma'}\rangle\nonumber\\
		=& \sum_{E_{\nu} < 0}v_{\nu,\mathbf{R}+\mathbf{r},\sigma}^{*}u_{\nu,\mathbf{R}-\mathbf{r},\sigma'},
\end{align}
which we decompose into singlet and triplet components
\begin{align}
	F^{s}(\mathbf{R}, \mathbf{r}) =& \frac{1}{2}\left(F(\mathbf{R}, \mathbf{r},\uparrow,\downarrow) - F(\mathbf{R}, \mathbf{r},\downarrow, \uparrow)\right),\nonumber\\
	F^{m=1}(\mathbf{R}, \mathbf{r}) =& F(\mathbf{R}, \mathbf{r}, \uparrow, \uparrow),\nonumber\\
	F^{m=0}(\mathbf{R}, \mathbf{r}) =& \frac{1}{2}\left(F(\mathbf{R}, \mathbf{r}, \uparrow, \downarrow) + F(\mathbf{R}, \mathbf{r}, \downarrow, \uparrow)\right),\nonumber\\
	F^{m=-1}(\mathbf{R}, \mathbf{r}) =& F(\mathbf{R}, \mathbf{r}, \downarrow, \downarrow).
\end{align}
Here $\mathbf{R}$ is the center-of-mass coordinate for a pair of electrons, while $\mathbf{r}$ and $-\mathbf{r}$ points to the two electrons.
The pair amplitude can further be classified as $s$- $p$- $d$-wave, and so forth, according to its angular dependence around the center-of-mass coordinate.
For this purpose the pair amplitude is further projected onto $e^{il\theta}$ to obtain the $s$-, $p_{+}$-, $p_{-}$-, $d_{+}$-, and $d_{-}$-wave pair amplitudes for $l = 0, 1, -1, 2, -2$, respectively.
In general, there is a certain degree of freedom in how to perform this projection, because the pair amplitude also has a radial dependence.
In practice, however, the pair amplitude is only expected to be sizable within a few coherence lengths, and it is therefore only the components of the pair amplitude which corresponds to small $\mathbf{r}$ that is of interest.
Focusing on the pair amplitude components for minimal $\mathbf{r}$ (without becoming trivially zero) for the various orbital moments, the appropriate projectors are
\begin{align}
	\label{Equation:Projectors}
	P_{s} = & \delta_{\mathbf{r}\mathbf{0}},\nonumber\\
	P_{s_{ext}} = & \frac{1}{4}\sum_{\mathbf{a}\in A}\delta_{\mathbf{r}\mathbf{a}},\nonumber\\
	P_{p_{\pm}} = & \frac{1}{4}\sum_{\mathbf{a}\in A}\delta_{\mathbf{r}\mathbf{a}}e^{\mp i\theta(\mathbf{r})},\nonumber\\
	P_{d_{\pm}} = & \frac{1}{8}\sum_{\mathbf{b}\in B}\delta_{\mathbf{r}\mathbf{b}}e^{\mp i2\theta(\mathbf{r})},
\end{align}
with
$	A = \{(1,0), (0,1), (-1,0), (0,-1)\}$,  
$	B = A\cup\{(1,1), (-1,1), (-1,-1), (1, -1)\}$,
and $\theta(\mathbf{r})$ the angular coordinate of $\mathbf{r}$.
The orbital wave functions onto which the pair amplitude is projected are displayed in Fig.~\ref{Figure:Projections}.
The projected pair amplitudes can now be written as
\begin{align}
	\label{Equation:Pair_function_orbital_and_spin}
	F_{O}^{S}(\mathbf{R}) =& \sum_{\mathbf{r}}P_{O}F^{S}(\mathbf{R}, \mathbf{r}),
\end{align}
where we study $O \in \{s, s_{ext}, p_{\pm}, d_{\pm}\}$ and $S \in \{s, 1, 0, -1\}$, while $\mathbf{r}$ runs over all possible lattice vectors.

\subsection{Even-frequency pairing}
\begin{figure*}
	\includegraphics[width=490pt]{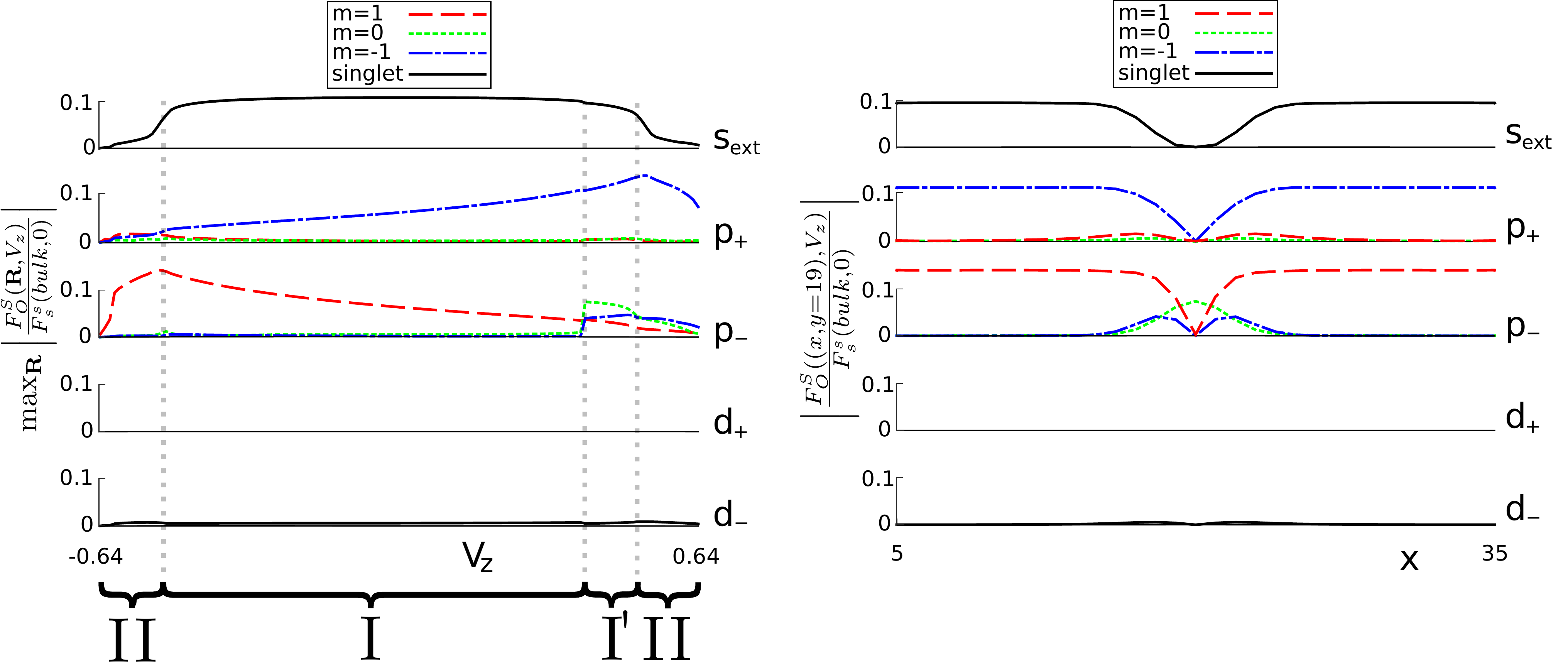}
	\caption{(Color online). Even frequency pair amplitudes.
		[Left] maximum absolute value of the pair amplitudes as a function of $V_z$.
		[Right] representative profiles for of the pair amplitudes along the $x$-axis, through the vortex core.
		All profiles are plotted for $V_z = 0.41$, just to the right of the transition from I to I', except the $m=1$ profiles where $V_z = 		-0.5$, but all pair amplitudes retain their profile characteristics as $V_z$ is varied.
		All values are normalized by the bulk value of the spin-singlet $s$-wave pair amplitude at zero Zeeman field, $F^{s}_s(bulk, 0)$.
		}
	\label{Figure:Pair_functions_even}
\end{figure*}

In Fig.~\ref{Figure:Pair_functions_even} we plot both the maximum value as a function of the Zeeman field, as well as representative real space profiles of the pair amplitudes.
All pair amplitudes are normalized by the bulk value of the conventional $s$-wave (singlet) pair amplitude at zero Zeeman field.
The conventional $s$-wave pair amplitude has been excluded from the plots as it is directly given by the relation $|\Delta(\mathbf{R})| = |-V_{sc}F(\mathbf{R}, 0, \downarrow, \uparrow)|$.
From the real space profiles it is clear that three types of behavior can be identified.
Namely, the pair amplitudes can be seen as originating in the bulk, vortex core, or region surrounding the vortex core, which here will be referred to as the pre-core region.
All pair amplitudes have their maximum value in one of these three regions, and decays to zero in the other two.
It is also clear that the three most important pair amplitudes in the bulk, apart from the $s$-wave (singlet) pair amplitude, are the extended $s$-wave (singlet), $p_{+}$ ($m = -1$), and $p_{-}$ ($m = 1$) components.
These all have a total angular momentum $z$-axis project $J_z = 0$, as expected if angular momentum is to be conserved.
The appearance of extended $s$-wave (singlet) is a direct consequence of conventional $s$-wave superconductivity; it only gives further details about the size of the pair amplitude at a finite radius $\mathbf{r}$.
The $p$-wave pair amplitudes in the bulk appear because of the Rashba spin-orbit interaction and is in agreement with previous results.\cite{PhysRevB.81.125318}
Our results do however provide further details showing that the $p_{+}$ ($m = -1$) component is largest when the Zeeman field is positive, while $p_{-}$ ($m = 1$) dominates for a negative Zeeman field.
This can be understood since essentially all electron levels are occupied in a lightly hole-doped semiconductor.
Thus the band crossing the Fermi level is the band which is pushed up in energy by the Zeeman field and contains predominantly spins anti-aligned with the Zeeman field.
This is also the band that is gapped by superconductivity, explaining the overweight of superconducting pairing among spins anti-aligned with the Zeeman field.
We note that the argument would be reversed for a lightly electron-doped semiconductor.

Next, we note that the existence of two $p_{-}$ states, in the vortex core ($m = 0$) and in the pre-core region ($m = -1$), are strongly correlated with the onset of the wide vortex core region I'.
For negative $V_z$, for which no vortex core widening occurs, we already have $p_{-}$ ($m = 1$) preferred in the bulk.
The rotation direction of the $p_{-}$ orbital part agrees with the rotation direction of the vortex.
For positive $V_z$, on the other hand, the $p_{+}$ ($m = -1$) state is preferred in the bulk but it has an orbital motion directed opposite to that of the vortex. This appears to lead to a widening of the vortex core.
Finally, we note that in the bulk and vortex core the total angular momentum is $J_z = 0$ and $J_z = -1$, respectively.
The angular momentum $J_z = -1$ of the vortex core pairing can be explained by the vortex winding $n = -1$ being absorbed into the orbital part of the pair amplitude.\cite{JPhysCSolidStatePhys.21.L215, JPhysCondensMatter.1.277, NewJPhys.11.075008, PhysRevB.84.064530, PhysRevB.88.104506}
On the other hand, the pre-core pair amplitudes respect neither the bulk nor the vortex core angular momentum conservation rule.

\subsection{Odd-frequency pairing}
Having described the behavior of the regular pair amplitudes we now turn to an investigation of odd-frequency pair amplitudes.
The possibility of odd-frequency pairing arise when $\mathbf{r}$ is considered to not only be a coordinate in space, but also in time, which we denote by $\mathbf{\tilde{r}} = (\mathbf{r}, t/2)$.
The ordinary requirement on the pair amplitude to be odd under the simultaneous interchange of position and spin $\langle c_{\mathbf{R}+\mathbf{\tilde{r}},\sigma}c_{\mathbf{R}-\mathbf{\tilde{r}},\sigma'}\rangle = -\langle c_{\mathbf{R}-\mathbf{\tilde{r}},\sigma'}c_{\mathbf{R}+\mathbf{\tilde{r}},\sigma}\rangle$ is then transformed to the requirement that the pair amplitude is odd under the simultaneous interchange of position, spin, and time (frequency).
The pair amplitudes investigated so far are all even in frequency, as only the even frequency components can be non-zero for $t = 0$.
To identify pair amplitudes which are odd in time, we have to consider the derivative of the pair amplitude with respect to time\cite{NewJPhys.11.065005, PhysRevB.86.144506}
\begin{align}
	\frac{dF(\mathbf{R}, \mathbf{\tilde{r}}, \sigma, \sigma')}{dt} =& \frac{d}{dt}\langle c_{\mathbf{R}+\mathbf{\tilde{r}},\sigma}c_{\mathbf{R}-\mathbf{\tilde{r}},\sigma'}\rangle\nonumber\\
		=& \frac{d}{dt}\sum_{E_{\nu} < 0}v_{\nu,\mathbf{R}+\mathbf{r},\sigma}^{*}u_{\mathbf{R}-\mathbf{r},\sigma'}e^{-\frac{i}{\hbar}E_{\nu}t}.
\end{align}
Preforming the decomposition of the total pair amplitude into its spin and orbital components as outlined in Eq.~(\ref{Equation:Pair_function}-\ref{Equation:Pair_function_orbital_and_spin}) allows us to also consider the amplitudes $dF_{O}^{S}(\mathbf{R}, t)/dt$.

\begin{figure*}
	\includegraphics[width=490pt]{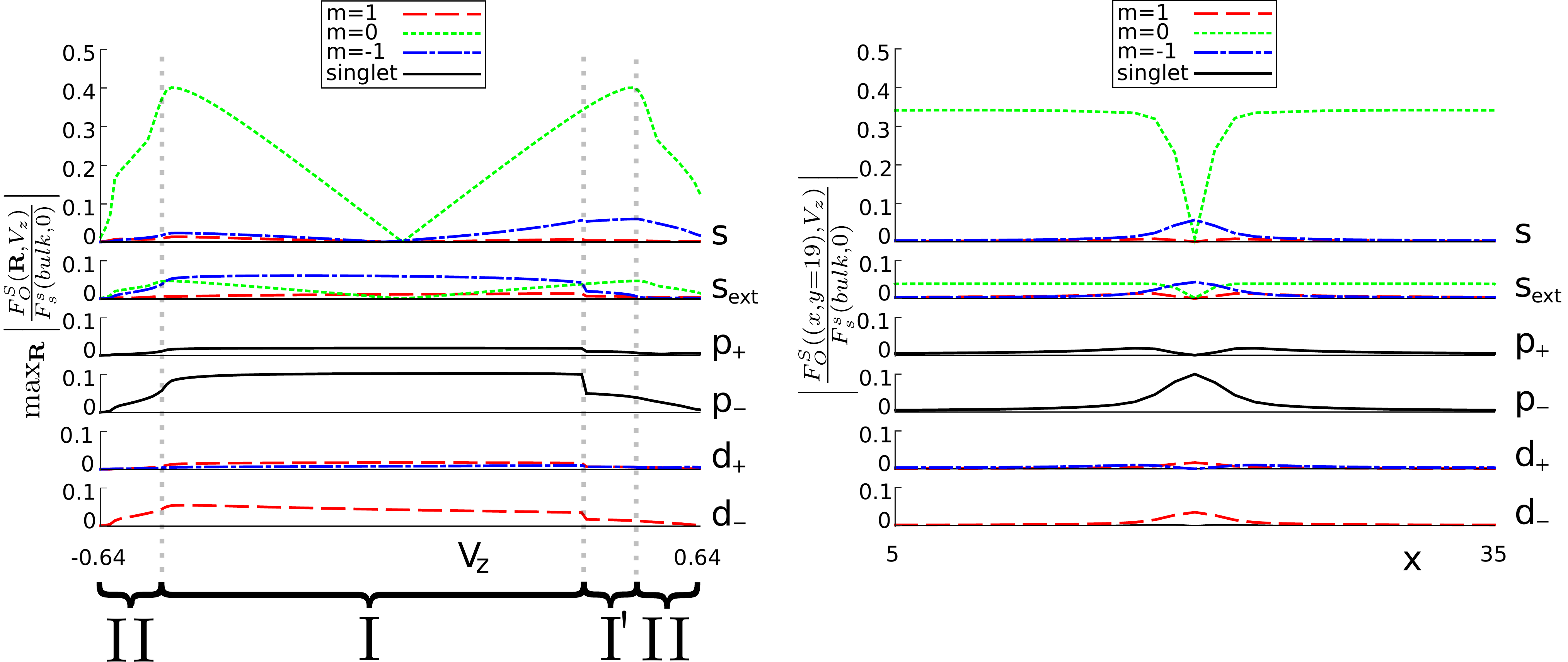}
	\caption{(Color online). Same as Fig.~\ref{Figure:Pair_functions_even}, but for odd frequency.
		All real-space profiles are for $V_z = 0.38$.
	}
	\label{Figure:Pair_functions_odd}
\end{figure*}

In Fig.~\ref{Figure:Pair_functions_odd} the odd frequency equivalent of Fig.~\ref{Figure:Pair_functions_even} is displayed.
Most notable is the appearance of an $s$-wave ($m = 0$) pair amplitude in the bulk when the Zeeman field is non-zero.
Likewise, the extended $s$-wave ($m = 0$) component is found to have a similar behavior.
All the pair amplitudes associated with the bulk have $J_z = 0$.

Next, we note four non-zero pair amplitudes in the core.
These are the $s$ ($m = -1$), extended $s$ ($m = -1$), $p_{-}$ (singlet), and $d_{-}$ ($m = 1$), which all have $J_z = -1$.
The existence of these can once again be understood as a consequence of the vortex winding being rotated into the pair orbital part.
However, although small, an anomalous $d_{+}$ ($m = 1$) component also appear in the core, violating the otherwise seemingly perfect agreement with vortex core pair amplitudes having $J_z = -1$.
This is due to the fact that each site has four nearest neighbors, while four of the sites included in the projection onto the $d$-wave pair functions are second-nearest neighbors.
On the nearest neighbor sites $e^{i2\theta(\mathbf{r})} = e^{-i2\theta(\mathbf{r})}$, so with respect to these site alone the angular momentum can only be defined modulo 4, and consequently $J_z = 3\textrm{\;mod\;}4 = -1$.
This is an example of how a radial variation in the pair amplitude, in this case the $d_{-}$ ($m=1$) component, can give rise to seemingly angular momentum violating components when calculated on a discrete lattice.
Unfortunately it is impossible to define a completely satisfactory projection procedure on a discrete lattice.
Finally, we note the existence of the pre-core type pair amplitudes $s$ ($m = 1$), extended $s$ ($m = 1$), $p_{+}$ (singlet), and $d_{+}$ ($m = -1$).
These, just as the pre-core type even-frequency $p_{\pm}$, all have $J_z$ satisfying neither $J_z = 0$ nor $J_z = -1$.
They do, however, all have in common that they either have a bulk or core counter-part differing only in spin or orbital rotation direction.
We therefore interpret these as being secondary in nature, being induced from their bulk and core counter-parts.

Before ending the discussion of pair amplitudes we also note that although the odd frequency $s$-wave ($m = -1$), as well as even frequency $p_{+}$ ($m = -1$) and $p_{-}$ ($m = 1$) pair amplitudes all increase as function of the Zeeman field towards the non-trivial phase, they contain no specific signature of the non-trivial phases themselves.
We put this in contrast to recent results relating the existence of odd-frequency $s$-wave pair amplitudes with the appearance of Majorana fermions in 1D.\cite{arXiv.1410.1245}
Likewise, the various pair amplitudes which appear in the vortex core provides little clue to the topological phase or existence of Majorana fermions, but appear in the whole phase diagram.
On the other hand, several of the pair amplitudes changes abruptly at the $I \leftrightarrow I'$ transition, thereby reflecting the sudden widening of the vortex core.
A summary of the described pair amplitudes is provided in Table \ref{Table:Pair_functions}.
%
\begin{table}[th!]
\begin{tabular}{c | c | c | c | c | c}
Orbital		& Frequency	& Spin		& $J_z$	& Origin & Max (\%)\\
\hline
\hline
$p_{+}$		& even		& $m = 1$	& $2$		& pre-core		& 2\\
$p_{+}$		& even		& $m = 0$	& $1$		& pre-core		& 1\\
$s_{ext}$	& even		& singlet	& $0$		& bulk			& 10\\
$p_{+}$		& even		& $m = -1$	& $0$		& bulk			& 14\\
$p_{-}$		& even		& $m = 1$	& $0$		& bulk			& 14\\
$p_{-}$		& even		& $m = 0$	& $-1$		& core			& 7\\
$p_{-}$		& even		& $m = -1$	& $-2$		& pre-core		& 5\\
$d_{-}$		& even		& singlet	& $-2$		& pre-core		& 1\\
\hline
$d_{+}$		& odd		& $m = 1$	& $3$		& core			& 2\\
$s$			& odd		& $m = 1$	& $1$		& pre-core		& 1\\
$s_{ext}$	& odd		& $m = 1$	& $1$		& pre-core		& 1\\
$p_{+}$		& odd		& singlet	& $1$		& pre-core		& 2\\
$d_{+}$		& odd		& $m = -1$	& $1$		& pre-core		& 1\\
$s$			& odd		& $m = 0$	& $0$		& bulk			& 41\\
$s_{ext}$	& odd		& $m = 0$	& 0		& bulk			& 5\\
$s$			& odd		& $m = -1$	& $-1$		& core			& 6\\
$s_{ext}$	& odd		& $m = -1$	& $-1$		& (pre-)core		& 6\\
$p_{-}$		& odd		& singlet	& $-1$		& core			& 10\\
$d_{-}$		& odd		& $m = 1$	& $-1$		& core			& 5\\
\hline
\end{tabular}
\caption{Even and odd pair amplitudes, ordered according to their $J_z$ quantum number.
	$J_z = 0$ and $J_z = -1$ pairing is seen to originate in the bulk and core, respectively, while other pair amplitudes only appears in the pre-core region.
	The $d_{+}$, odd frequency, $m = 1$ pair amplitude is anomalous in that it originates in the core region although it has $J_z = 3$.
	As explained in the text, it is an artifact of the pair amplitudes being calculated on a lattice.
	Likewise, the $s_{ext}$, odd frequency, $m = -1$ pair amplitude is anomalous in the wide vortex core region, as it there originates in the pre-core region rather than in the core, in spite of having $J_z = -1$.
	The maximum relative strength as compared to the $s$-wave, even frequency, spin-singlet pair amplitude at zero Zeeman field is tabulated at the right.
}
	\label{Table:Pair_functions}
\end{table}

\section{Summary}
We have in this work investigated the local density of states (LDOS), band structure, and superconducting pair amplitude for signatures of the non-trivial topological phase and Majorana fermions in vortex cores in spin-orbit coupled semiconductor-superconductor heterojunctions.
A necessary indicator of a zero-energy vortex core state being a Majorana fermion has been identified to be a momentum distribution centered at a finite radius away from the high symmetry point $\mathbf{k} = (\pi, \pi)$, [$\mathbf{k} = (0, 0)$ in the case of a lightly electron-doped semiconductor]. Moreover, the vortex Majorana fermion and finite-energy Caroli-Matricon-de Gennes vortex states are found to be well separated in energy and in the topological phase they together form a characteristic x-shape structure in the subgap LDOS when scanning through the vortex core. The Majorana mode is very well-localized in the center of the core, while the finite-energy states disperse further out from the center, although the x-shape structure is still centered at the core center.

Furthermore, we show that a clear signature in the spectral function of the topological phase itself is the Mexican hat shaped band structure, which also gives rise to double band edges, very clearly visible in the DOS due to their high concentration of states.
These double band edges also give rise to the existence of a second class of vortex core states, distinct from the ordinary Caroli-Matricon-de Gennes vortex core states and the Majorana fermion.
These vortex states appear beyond the superconducting gap and forces locally larger band gap in the vortex core region, both features that are experimentally measurable.

Finally, we have also investigated the superconducting pair amplitude, showing that multiple pair amplitudes with total $J_z = 0$, both even and odd in frequency, develops in the bulk because of the finite spin-orbit interaction and magnetic field. In the vortex core we instead find pair amplitudes which have a total $J_z = -1$, where the vortex core momentum ($-1$) has been rotated into the orbital part of the pair amplitudes. Despite multiple unconventional pairing amplitudes developing in the core, we find no amplitude that signals the onset of non-trivial topological order. Specifically, the appearance of a Majorana fermion does not imply any noticeable increase in the odd-frequency pairing.
However, we find a strong correlation between the transition from narrow to a wide vortex core and the development of even-frequency $p_{-}$ components.
In summary, these results provide multiple specific characteristics for the non-trivial topological phase and its vortex Majorana fermion in spin-orbit coupled semiconductor-superconductor heterostructures. These distinct indicators provide both added physical understanding of the topological phenomena in these heterostructures and are in many cases directly experimentally measurable.

\acknowledgments
We are grateful to A.~Bouhon, A.~Balatsky, and S.~Pershoguba for useful discussions and the Swedish Research Council (Vetenskapsr\aa det), the G\"oran Gustafsson Foundation and the Swedish Foundation for Strategic Research (SSF) for financial support.
The computations were performed on resources provided by SNIC through Uppsala Multidiciplinary Center for Advanced Computational Science (UPPMAX) under project p2012124 and snic2014-1-280.

\end{document}